\documentclass[12pt]{iopart}

\usepackage[dvips]{graphicx}
\def\lapprox{\hbox{\lower .8ex\hbox{$\,\buildrel < \over\sim\,$}}}
\def\gapprox{\hbox{\lower .8ex\hbox{$\,\buildrel > \over\sim\,$}}}

\begin{document}

\title[ ]{Dark energy,
gravitation and supernovae}

\author{Pilar Ruiz--Lapuente}

\address{Department of Astronomy and CER for Astrophysics, Particle
 Physics and Cosmology, University of Barcelona}
\ead{pilar@am.ub.es}
\begin{abstract}
The discovery of the 
acceleration of the rate of expansion of the Universe fosters  
new explorations of the behavior of gravitation theories in the cosmological
context. Either the GR framework is valid but a cosmic component
with a negative equation of state is dominating the energy--matter contents
or the Universe is better described at large by a theory that departs from GR. 
In this review we address theoretical alternatives 
that have been explored through supernovae. 

\end{abstract}

\maketitle

\noindent{\bf Contents} 

\smallskip

\noindent
1. Introduction 

\smallskip 

\noindent
2. Gravity and extra dimensions 

\smallskip 

\noindent
3. The dark energy field and scalar--tensor gravity 

\smallskip 

\noindent
4. The cosmological constant? 

\smallskip 

\noindent
5. The d$_{L}$ test from high--z SNe Ia 

\smallskip 

\noindent
6. Testing the adequacy of the FRW metric 

\smallskip 

\noindent
7. Next decade experiments 

\smallskip 

\noindent
8. Complementarity

\smallskip 

\noindent
9. Conclusion

\vfill\eject

\section{Introduction}

The discovery of the acceleration of the expansion of the universe 
\cite{riess98,perlmutter99}
has revived questions concerning the domain of validity of Einstein's General
Relativity. At present, the data gathered on the expansion
rate do not disclose whether the acceleration is due to a component
formally equivalent to the cosmological constant introduced by Einstein 
\cite{einstein}, or
whether we are finding the effective behavior of a  theory of a wider scope
whose low energy limit slightly departs from GR.

\noindent
The bare Einstein's equations (without cosmological constant) 
for a Universe with a FRW metric and dust--like matter 
 imply a continuous deceleration of the expansion rate. However, a  
Universe containing a fluid with an equation of state $p=w \rho$ with index 
$w< -1/3$ overcomes the deceleration when the density of this fluid
dominates over that of the dust--like matter. In this context, 
the cosmological constant, if it is 
positive and added to the equations, balances the deceleration by  
acting as a fluid with an equation of state $p= - \rho$. 

\noindent
The cosmological constant was originally in the
metric part of Einstein's equations, and alternative metric
theories of gravity provide a wealth of terms in that side. 
When describing a homogeneous and isotropic universe
within a FRW model, it is frequent to come as close as possible 
to the standard 
Friedmann--Einstein equation for the expansion rate by bringing terms to
the right--hand side within 
 the so--called  ``effective index'' $w_{\rm eff}$ of the cosmic fluid whose 
equation of state is  $p= w_{\rm eff} \rho$.

\noindent
In the field of gravitation,
ideas that appeared in the classical 
domain several decades ago, are now being brought back within the
different context of quantum gravity theories. 
Scalar--tensor theories of gravitation are being reconsidered.  
Scalar fields coupled to gravity were proposed as a way to  
incorporate Mach inertial induction. They gave rise to 
the Brans--Dicke theory of gravitation \cite{bransdicke}, which is well 
restricted by solar system experiments\cite{will93}. Superstring theory and 
M--theory introduce a number of quantum scalar fields, the dilaton, the
radion, the moduli \cite{green}. They can play a role in the late
acceleration of the Universe. 

\noindent
On the other hand, brane worlds inspired by M--theory provide
modifications to gravity and 
new equations of cosmic expansion \cite{arkani-hamed}.
 In the brane world context, gravity  
lives in the bulk of 3+1+d dimensions while the rest of the interactions
are confined to the 3+1
brane. Some braneworld proposals do not explain the observed cosmic expansion. 
However, it was realized \cite{dvali, deffayet} that 
it is possible to obtain late--time acceleration with one
extra dimension. This possibility is interesting as it accounts for 
the observed cosmic expansion without the need of 
a cosmological constant term. Acceleration results from the induced gravity
term in the brane. The number and size of additional dimensions has been 
explored in a wide range of proposals. The existence of 
additional dimensions and their size are constrained by tests of the 
gravity law \cite{adelberger03,adelberger05,adelberger06}. 

\noindent
At the time when gravity theories of the scalar--tensor type and
others were examined, the solar system tests were the most restrictive ones. 
Nowadays, the theories have to confront the
cosmological observations. The theories have to allow  Big Bang 
nucleosynthesis \cite{steigman}.  
They have to fullfil the constraints from the growth 
of density perturbations and give correct predictions for the evolution
 of the growth
factor. Such growth function of density perturbations requires careful
evaluation for several modified gravity scenarios 
 \cite{koyama,maartens}. Ultimately, gravity theories 
 have to be consistent with the CMB data and with the expansion rate 
of the Universe provided by supernovae.

\noindent
Supernovae enable to measure the rate of the expansion of the Universe 
along redshift z. They do so by tracing the luminosity distance along z 
for a standard candle. If the Universe is flat, as inferred from WMAP 
\cite{spergel}, the $d_{L}$ relation is simply: 

\begin{equation}
d_{L}(z) = c H_{0}^{-1}(1 + z) \int_{0}^{z} {dz'\over{H(z')}}
\end{equation}

\noindent
while in the general case, 

\begin{equation}
d_{L}(z) = c H_{0}^{-1} (1 + z) |\,\Omega_{K}|^{-1/2}\, {\rm sinn}
\left(|\,\Omega_{K}|^{1/2} \int_{0}^{z} {{H_{0} dz'}\over{H(z')}}\right)
\end{equation}

\noindent
where ${\rm sinn} (x)$ $=$ 
${\rm sin}(x)$, $x$, or ${\rm sinh}(x)$ for closed, flat, 
and open models respectively, where the curvature parameter $\Omega_{K}$, 
is defined as $\Omega_{K} =  \Omega_{T} - 1$.

\noindent
$H(z)$ is the Hubble parameter. For a FRW universe containing 
matter and a general form of matter--energy $X$  with index of
the equation of state $w_{X}$, $H(z)$ is given by:  

\smallskip

\begin{equation}
H^{2}(z) = H_{0}^{2} \left\{ \Omega_{K}(1 + z)^{2} + \Omega_{M}(1 + z)^{3}
+ \Omega_{X}(1 + z)^{3(1+w_{X})}\right\}
\end{equation}

\noindent
where $H_0$ is the present value of the Hubble parameter and 

\begin{equation}
\Omega_{K} \equiv {{-k}\over{H_{0}^{2}a_{0}^{2}}}
\end{equation}

\noindent
where k accounts for the geometry and $a_{0}$ is the present value
of the scale factor. 

\begin{equation}
\Omega_{M} = {{8\pi G}\over{3H_{0}^{2}}}\rho_{M}
\end{equation}

\noindent
 $\Omega_{X}$ is the density parameter for dark energy. 

\smallskip

\noindent
When $w(z)$ varies along z, $H(z)$ can be expressed as:

\begin{eqnarray}
H(z)^{2} = H_{0}^{2} \left\{\Omega_{K}(1 + z)^{2} + \Omega_{M}(1 + z)^{3}
+ \Omega_{X}\, {\rm exp}\, \left[3\int_{0}^{z} {{1 + w(z')}\over{1 + z'}}dz'
\right]\right\}
\end{eqnarray}

\noindent 
One sees from (1--2), that $d_{L}$ contains $w(z)$ through a double integral.
This leads to limitations in recovering $w(z)$ if 
$d_{L}$ data along z are scarce and with large errorbars. 

\noindent
Data are often given as $m(z)$ which relates to  $d_{L}$(z) as:

\begin{eqnarray}
m(z) & = 5\; {\rm log} d_{L}(z)  + [M + 25 - 5\; {\rm log} (H_{0})]
\end{eqnarray}

\noindent
The quantity within brackets is 
${\cal M} \equiv M - 5\, {\rm log}\, H_{0}
+ 25$, the ``Hubble-constant-free'' peak absolute magnitude of a
supernova. The data are given sometimes in the form of distance moduli: 

\begin{equation}
\mu = 5\ {\rm log}\ d_{L} + 25
\end{equation}

\noindent
The measurement of $w(z)$ through $d_{L} (z)$ is the  standard 
approach, though possibilities of measuring directly $H(z)$ have
been investigated (for a review see \cite{sahni06}).  

\noindent
This Hubble law in (3) 
is derived from the Einstein-Hilbert action. We will review 
other possibilities related to expansion histories arising from
modified gravity theories and effective theories which contain
scalar fields accelerating the cosmic expansion. In the present 
overview we concentrate on the empirical evaluation of those theories.
Some of the models to be explored here might contain a ghost problem or have
a subclass amongst them that present quantum instabilities. We refer for
this topic to recent updates in 
\cite{calcagni,defelice,gorbunov,padilla,trodden}.

\section{Gravity and extra dimensions}

The idea of allowing gravity to live in extra dimensions 
arose to explain the hierarchy problem and constitutes one of 
the major challenges to the classical picture of a 4--dimensional
description of gravity \cite{arkani-hamed}. 
 The Planck scale of 10$^{19}$ GeV,
relevant for gravitation, is much larger than the electroweak scale,
of about 1 TeV of the standard model of particle physics. According to the 
proposal by Arkani-Hamed, Dimopoulos and Dvali \cite{arkani-hamed},
 the Planck scale
is not a fundamental scale, but its enormity is simply a consequence 
of the large size of new extra dimensions. While gravitons can propagate
in the new extra dimensions, the standard model
fields are confined to our submanifold of 3 $+$ 1 dimensions, the brane,
which is embedded in a higher dimensional bulk. 

\noindent
In this scenario, the Planck scale in the bulk can be 
much smaller than in the $4$--dimensional brane as 

\begin{equation}
M_{\rm P,4}^{2} = R^{n}M_{\rm P,n+4}^{\rm n+2} 
\end{equation}

\noindent
The existence of extra dimensions in which gravity propagates produces a 
transition in the gravitation law at the compactification radius, proposed
to be of the submillimetre scale.

\smallskip

\noindent
In the cosmological context, this idea has some interesting
consequences since it predicts a departure from GR. As soon as
the cosmological model was implemented for $n=1$ extra
dimensions \cite{lukas,binetruy99,deffayet,dvali}, it was noted that 
the brane cosmology leads to Friedmann--like equations very different
from the standard ones \cite{binetruy99}. 
In the 5--dimensional model by Dvali, Gabadadze and Porrati (DGP) whose 
cosmology is explored 
by Deffayet and collaborators 
in \cite{deffayet}, 
the competition between the 
five--dimensional term of the Einstein--Hilbert action and the 
four--dimensional term induced in the brane leads to a
modified Friedmann equation which simulates an acceleration 
term without the need to include a cosmological constant.
The transition from the gravity law  $r^{-2}$ to the modified 
one is characterized by a scale $r_{c}$, the so--called 
cross--over scale defined as \cite{deffayet}: 

\begin{equation}
{r_{c}} \; \equiv \; {{M_{\rm P}^{2}} \over{2 M_{(5)}^{3}}}
\end{equation}

\noindent
 At large distances, the 
five dimensional term takes over and gravity gets weaker \cite{deffayet}. 
If the parameter $r_{c}$ is chosen to be of the order of 
H$_{0}$$^{-1}$ $\sim$ 10$^{29}$ mm, this is a choice for  
$M_{(5)}$ $\sim$ 10--100 MeV. At short distances $r << r_{c}$, Newton's law
is modified by a logarithmic repulsion. At large distances $r >> r_{c}$, 
the modification involves a factor $r\over{r_{c}}$.  

\noindent
The Hubble expansion law 
shows significant departures from the standard Friedmann
expansion law:

\begin{equation}
H^{2}(z) = H_{0}^{2} \left\{ \Omega_{K}(1 + z)^{2} + \left( \sqrt{
\Omega_{r_{c}}} + \sqrt{\Omega_{r_{c}} + \sum_{\alpha} 
\Omega_{\alpha}(1 + z)^{3(1+w_{\alpha})}}\right)^{2}\right\}
\end{equation}

\noindent
where the sum over $\alpha$ stands for a sum over each component with
equation of state $w_{\alpha}$, $\Omega_{\alpha}$ being the density
parameter:

\begin{equation}
\Omega_{\alpha} \equiv \frac{\rho_{\alpha}^{0}}{3M_{\rm Pl}^{\;\;\;2} H_{0}^{2}
a_{0}^{3(1+w_{\alpha})}}
\end{equation}

\noindent
and

\begin{equation}
\Omega_{r_{c}} \equiv {1\over{4r_{c}^{2}H_{0}^{2}}}
\end{equation}

\noindent
In the matter--dominated epoch, the expansion equation is:

\begin{equation}
H^{2}(z) = H_{0}^{2} \left\{ \Omega_{K}(1 + z)^{2} + \left( \sqrt{
\Omega_{r_{c}}} + \sqrt{\Omega_{r_{c}} + 
\Omega_{M}(1 + z)^{3}}\right)^{2}\right\}
\end{equation}

\noindent
In the original 5D model in \cite{dvali}, $M_{5}$ was estimated to be 
$\sim$ 1 TeV, which corresponds to a cross--over scale $r_{c}$ of 10$^{15}$ cm.
Such $r_{c}$ led to incompatibilities with the solar system data.
When  $M_{5}$ is taken to be $M_{5} \lapprox $ 1 GeV, i.e. $r_{c} \gapprox$ 
10$^{25}$ cm, the conflict disappears.

High--z supernovae give useful constraints to the parameter  $\Omega_{r_{c}}$
in (10) \cite{fairbairn,majerotto}. In \cite{majerotto}, two samples of 
high--z supernovae are used in conjunction 
 with measurements from baryon oscillations \cite{eisenstein05} and the
 CMB shift parameter \cite{wang} to estimate the favored values for 
 $\Omega_{r_{c}}$ and $\Omega_{M}$.  $\Omega_{r_{c}}$ $=$ 0.125 for the 
 supernova data in the Riess sample from 2004 \cite{riess04} and 
  $\Omega_{r_{c}}$ $=$ 0.13 for the SNLS data \cite{astier06}. It is found
 in \cite{majerotto} that, for a flat Universe, this 5D modified gravity
 model gives a worse fit than GR plus cosmological constant.  
 Further supernova samples would reexamine the empirical compatibility
 of the DGP model and other modified gravity models involving more dimensions.

\noindent
Those models can be explored through a 
phenomenological description. The Hubble law  arising
from gravity with extra dimensions can be expressed by adding 
$H^{\alpha}$ terms
to the standard Friedmann equation \cite{dvaliturner}:

\begin{equation}
H^{2} - {{H^{\alpha}}\over{r_{c}^{2-\alpha}}} = {{8\pi G\rho_{M}} \over3}
\end{equation}

\noindent
where $r_{c}$ is the cross--over radius, i.e. the transition radius from a
regime to another in the gravity law, and $\alpha$ is the
parameter to be empirically determined. From redshift z$=$ 0 to z$=2$, 
the above Friedmann equation gives an effective index of the equation of state
\cite{dvaliturner}:

\begin{equation}
w_{\rm eff}  \approx  -1 + 0.3 \alpha
\end{equation}

\noindent
Such effective equation of state evolves too fast to be compatible with
present SNe Ia data, unless $\alpha$ is small. Present data
 favor  $\alpha$ close 
to 0 \cite{fairbairn}. 

\noindent
Though the most simple implementations lack empirical confirmation,  
room is left for several  models.  Cosmological SNe Ia are useful tests in
restricting 
the cross--over radius that relates to departures from GR.
 
\smallskip

\noindent
A different kind of model in which the brane is embedded in a 5D bulk
has been proposed by Randall and Sundrum 
\cite{randall-sundrum1,randall-sundrum2}. Contrary to the
previous approach, here the bulk has a $AdS_{5}$ geometry as opposed to
the flat geometry of the DGP model.

\noindent
In the original Randall--Sundrum (R--S)
model, the brane has tension $\sigma$, i.e. cosmological constant,
 as well as the
bulk $\Lambda_{5}$. The bulk cosmological constant is negative.
The resulting Hubble expansion in the brane 
is the result of the compensating effect of both terms.  
To make zero the effective cosmological constant in the brane,
$\Lambda_{4}$, a fine tuning
between the brane tension and the bulk cosmological constant is
required. The original proposal sets  
the effective cosmological constant, $\Lambda_{4}$, to zero. 
Generally, a tiny deviation from exact compensation \cite{binetruy0}
would lead to an effective cosmological constant at late times.  
Modifications of the original R--S proposal include in the brane, 
in addition to tension (cosmological constant),
 a matter--energy density $\rho$.
The total brane density $\rho_{B}$ is:

\begin{equation}
\rho_{B} = \sigma + \rho
\end{equation}

\noindent
This gives the modified Friedmann equation \cite{langlois}:

\begin{equation}
H^{2} = \left({{\kappa^{4}}\over36} \sigma^{2} - {1\over{l^{2}}}\right) 
+ {{\kappa^{4}}\over18} \sigma \rho + {{\kappa^{4}}\over36} \rho^{2} +
{{\cal C}\over{a^{4}}}
\end{equation}

\noindent
where $\kappa$ is the inverse Planck mass $M_{\rm P}^{-1}$ and 
 $l$ is related to $\Lambda_{5}$. 

\noindent
Generalizations of the Randall--Sundrum scenario in cosmology consider
the {\it  AdS$_{5}$} bulk geometry containing the fifth--dimensional
cosmological constant $\Lambda_{5}$, but explore arbitrary densities 
in the brane  and in the bulk.

\noindent
Explorations of  
braneworld ideas that allow a late acceleration of the Universe
can be found in \cite{braxbruck,easson,durrer}.

\section{The dark energy scalar field and scalar--tensor gravity} 

\subsection{Early developments} 

Before the development  of braneworld scenarios, 
dark energy as associated to a scalar field was considered in 
analogous way to research done within the context of inflation
 \cite{ratra88}. This 
theoretical possibility was entertained 
a decade before there was evidence of a non--zero cosmological
constant from supernovae and that the CMB results indicated a flat Universe. 
The favored $\Omega_{T} = 1$ value by inflation and the 
evidence that $\Omega_{M}$ $<$ 1 led to think about a possible $\Lambda$
term. In the recent expansion history a scalar field could play a role
in analogy with inflation, where a 
 scalar field dubbed the inflaton gives rise to the exponential
phase of early acceleration. The scalar field responsible for the
late acceleration, later named quintessence, is
proposed as a new massless scalar field which
induces a cosmological constant behavior at late times.
 
\noindent
In this original setting, the field is minimally coupled to
gravity, i.e., thus the inclusion of the field remains within the framework
of GR. The interaction of the field with the matter fields is set to
be negligible. 

\noindent
The scalar field responsible for dark energy can be identified
by recovering its potential, or equivalently its effective 
equation of state.  The presence of such scalar field $\phi$, would 
induce an action:

\begin{equation}
S = \int d^{4}x\sqrt{-g}\left[{1\over2}g^{\mu\nu}\partial_{\mu}\phi
\partial_{\nu}\phi - V(\phi)\right]
\end{equation}

\noindent
The field contributes to the stress--energy momentum tensor
with an effective mass density and pressure:

\begin{eqnarray}
\rho_{\phi}  &\; =\; & {1\over2}\dot\phi^{2} + V(\phi) \nonumber \\
p_{\phi} & \;=\; & {1\over2}\dot\phi^{2} - V(\phi)
\end{eqnarray}

\noindent
In this case, the effective value of the equation of state $w_{\rm eff}$
depends on the form of the potential 
$V(\phi)$ and it can evolve with time \cite{weller00}.

\begin{equation}
w_{\rm eff} = {{p_{\phi}}\over{\rho_{\phi}}} 
\end{equation}

\noindent
A search for general properties of this potential, in analogy to 
the search for the inflation potential, leads to suggestions for 
potentials with tracking behavior \cite{ratra88, wetter88, steinhard99, 
peebles03}. 
Several potentials have 
been proposed, from the original $V = \kappa/\phi^{\alpha}$ \cite{ratra88} 
to other recent proposals \cite{copeland06}.

\noindent
The dark energy scalar field that does not couple
with matter has the feature of avoiding effects
 on variations of constants and other 
effects subject to  experimental evidence \cite{uzan03,bertolami}.
There is already a reconstruction of what could be the potential 
of the minimally coupled field $V(\phi)$ obtained from the SNLS
SNe Ia \cite{cooray} following reconstruction ideas 
given in \cite{huterer99,huterer00}. Another reconstruction of 
$V(\phi)$ using SNe Ia from various collaborations and considering 
Pad\'e approximant expansions of the potential is found in \cite{sahlen}.
By now, many proposals have been discarded, since the equation 
of state between 
z$=0$ and z$=1.5$ evolves very slowly as shown by high--z SNe Ia data 
\cite{perlmutter99,knop03,riess04,riess06}.

\subsection{Scalar--tensor gravity}

One can still consider the case of a scalar field that is coupled
to other matter fields or to gravity. Some of these proposals come 
from the ballpark of superstring theory. 

\noindent
The most well--known scalar--tensor gravity proposal, i.e. Brans--Dicke 
theory, incorporated 
the coupling of a field to gravity to account for the Mach 
induction on Newton's constant. In Brans--Dicke gravity 
 \cite{brans05,bransdicke}, 
Newton's constant is replaced by a dynamical field 
$G$ leading to the generalized Friedmann equation: 

\begin{equation}
H^{2} = {{8\pi}\over3} {\omega\over{2\pi \phi^{2}}}
\left(\rho + {1\over2}\dot\phi^{2} 
+ V(\phi) - {3\over{2\omega}}H\phi\dot\phi\right)
\end{equation}

\noindent
The Brans--Dicke action  contains a $\phi^{2} R$ term:

\begin{equation}
S = \int \sqrt{-g}\;d^{4}x\left[\pm {1\over{8\omega}}\phi^{2}R\mp{1\over2}
g^{\mu\nu}\phi_{,\mu}\phi_{,\nu} - V(\phi) + {\cal L}_{\rm matter}\right]
\end{equation}

\noindent
The theory 
has been well tested \cite{will93}. Observationally, for other scalar--tensor 
theories, we aim at retrieving the  self--interaction potential of the scalar
$V(\phi)$. and the coupling function to matter and gravity 
$A(\phi)$ in the matter term of the action \cite{esposito-farese}:

\begin{equation}
S_{\rm matter}(\psi, A^{2}(\phi) g_{\mu\nu})
\end{equation}

\noindent
In the case of quintessence (minimally coupled field) $A(\phi) =1$. 
The present supernova data have explored  the potential for 
a minimally coupled field \cite{cooray}. In the non-minimally coupled case,
also referred to as coupled quintessence  or coupled dark energy
\cite{amendola99,amendola03}, the CMB data can put constraints on the strength
of scalar gravity as compared to ordinary tensorial gravity
 \cite{amendola99}. However, 
there are limitations to the recovery of the form of the coupling and the 
coupled dark energy potential from CMB observations  \cite{amendola99}. 
Supernovae 
together with observational constraints on the growth of density 
perturbations would allow to retrieve  $A(\phi)$ \cite{boisseau0,espol}. 

\noindent
We expand $A(\phi)$ in its derivatives $\alpha_{0}$, $\beta_{0}$ and 
higher orders: 

\begin{equation}
{\rm ln } \, A(\phi) \equiv \alpha_{0} (\phi -\phi_{0}) + 
{ 1\over2} \beta_{0} (\phi - \phi_{0})^{2} + 
{\cal O} (\phi - \phi_{0})^{3}
\end{equation}

\noindent
By exploring the empirical evidence it is found \cite{esposito-farese} that 
solar system tests and binary pulsar tests allow to impose 
precise bounds on the first and second derivatives of the matter--scalar
coupling function, while SNe Ia could ``a priori'' allow to reconstruct
the full shape of the function $A(\phi)$ or the higher order derivatives.
 It is argued in \cite{boisseau0,espol,periva}
 that the knowledge of the luminosity distance
 and the density fluctuations $\delta \rho / \rho$ as functions of 
redshift z is sufficient to reconstruct the potential $V(\phi)$ and 
the coupling function $A(\phi)$.

\noindent
These results can be generalized to proposed actions that can be
re--expressed as a theory of 
interacting matter fields in general relativity \cite{tourenc}.
Actions containing $e^{-\Phi} R$  are motivated from 
the lowest order effective action including the dilaton $\Phi$ 
\cite{gaspven}.
They can be brought by 
redefinition of the field to an Einstein frame in which the field is 
minimally coupled to the metric. 

\noindent
A number of interesting cases have been examined and have been  
expressed in terms of a scalar field which is minimally coupled 
to the metric, but nonminimally coupled to the other fields 
\cite{barrow88,barrow90,holdenwand}. The self--interaction potential 
 shows an explicit energy transfer between the scalar field and
the matter fields.

\noindent
Related to the previous exploration is that of {\it f(R)} gravity or 
 extended curvature gravity \cite{wands}, also known as
 ``c--essence'' when including an inverse power of R \cite{chiba1}.
Those modified gravity theories with a Lagrangian containing an
arbitrary function of $R$ are equivalent to a particular class of 
scalar--tensor theories of gravity. For instance, the  action

\begin{equation}
S = {1\over{2\kappa^{2}}} \int d^{4}x\; \sqrt{-g}\left(R - 
{{\mu^{4}}\over{R}}\right) + S_{\rm matter}(g_{\mu\nu})
\end{equation}

\noindent
 is found to be equivalent to \cite{chiba1}:

\begin{equation}
S = {1\over{2\kappa^{2}}} \int d^{4}x\; \sqrt{-g}\left(\left(1 +
  {{\mu^{4}}\over{\phi^{2}}}\right)R - {{2\mu^{4}}\over{\phi}}\right) + 
S_{\rm matter}(g_{\mu\nu})
\end{equation}

\noindent
The equivalence is generalized to an arbitrary function of
R \cite{tourenc,wands}:

\begin{equation}
S = {1\over{2\kappa^{2}}} \int d^{4}x\; \sqrt{-g}\; F(R) + 
S_{\rm matter}(g_{\mu\nu})
\end{equation}

\noindent
with an equivalent action \cite{tourenc,wands}:

\begin{equation}
S = {1\over{2\kappa^{2}}} \int d^{4}x\; \sqrt{-g}\; (F(\phi) + 
F'(\phi)(R - \phi)) + S_{\rm matter}(g_{\mu\nu})
\end{equation}

\noindent
where $F'(\phi) = dF/d\phi$.

\noindent
The introduction of a canonical scalar field $\varphi$ such that $F'(\phi) =
{\rm exp}(\sqrt{2/3}\; \kappa \varphi)$, allows to rewrite the action
with F(R) as:

\begin{eqnarray}
S = \int d^{4}x \sqrt{-g_{E}}\left({1\over{2\kappa^{2}}}R_{E} -
{1\over2}(\nabla\varphi)^{2} - V(\varphi)\right) + S_{\rm matter}
(g_{\mu\nu}^{E}/F'(\phi(\varphi)), \nonumber \\
V(\varphi) = (\phi(\varphi)F'(\phi(\varphi)) - F(\phi(\varphi)))/2\kappa^{2}
F'(\phi(\varphi))^{2} 
\end{eqnarray}

\noindent
The gravity described by $g_{\mu\nu}^{E}$ is the Einstein--scalar system and
$g_{\mu\nu}$ ($= g_{\mu\nu}^{E}/F'(\phi)$). 
$g_{\mu\nu}$ is a metric of scalar--tensor gravity which is, in this
case, subject to constraints coming from solar system experiments. Again, a
classification of when one recovers an action equivalent to 
the Einstein--Hilbert case and when one recovers a Brans--Dicke--type
action in {\it f(R)} gravity is given in \cite{tourenc}. When considering 
a general function of R, it is not always possible to obtain a description 
equivalent to Einstein GR plus additional fields minimally coupled
 to the metric
but coupled to other fields. When this is possible, we have in the matter term 
the coupling between the matter fields
$\psi$ and the scalar field $\phi$. 
Examples of {\it f(R)} casted in terms of fields can be found 
in \cite{barrow88,holdenwand,star00,chiba1}.

\noindent
For the Starobinsky model \cite{star00} with 
$F(R) = R + R^{2}/M^{2}$ with $M \sim 10^{12}$ GeV,
for instance, the effective potential can be rewritten, in terms of the scalar
field $\varphi$, as

\begin{equation}
V(\varphi) = {{M^{2}e^{-2\sqrt{2/3}\;\kappa\varphi}}\over{8\kappa^{2}}}
(e^{\sqrt{2/3}\;\kappa\varphi} -1)^{2}
\end{equation}

\noindent
Actions with generalized inverse powers of $R$ have been investigate as well
\cite{cappozziello,carroll05,mena06,sotiriou}.

\noindent 
The supernovae gathered up to now are  compatible with 
some  {\it f(R)} gravity models and incompatible with others
 \cite{cappozziello,mena06,amendolpol}.

\noindent
More general scalar--tensor theories than the ones here defined,
such as those where the potential depends on several scalar fields 
$V(\phi^{n})$ and  matter--scalar coupling  $A(\phi^{n})$, 
give rise to a wider phenomenology.

\noindent
In the superstring scenario, 
nonminimal coupling of the scalar fields to the spacetime curvature 
leads to  effective actions with similarities to scalar--tensor 
proposals. 
Some scalars present in the effective action arise from
compactification to four dimensions. Those are structure moduli which do not
couple directly with the spacetime curvature tensor. 
One modulus field $\sigma$ 
associated with the overall size of the internal compactificaton 
couples to Einstein gravity via Riemann curvature invariants, such 
as the Gauss--Bonnet invariant \cite{antonrizos}.
If one considers the
Gauss--Bonnet term, at least an extra
function, the coupling to the Gauss--Bonnet term, $W(\sigma)$ needs 
to be determined.

\noindent
In the following, we review attempts to include such enlarged action and
 their test
with cosmological d$_{L}$ data. 

\subsubsection{Effective low energy action with R$_{GB}$ and other terms} 

The Gauss--Bonnet term can be found in the effective
low energy string Lagrangian and in brane theories \cite{antoniadis, 
antonrizos, neupane,langlois,gross}. 
The case of the simplest action containing the Gauss-Bonnet term
can be written as: 

\begin{eqnarray}
S = &  \int \sqrt{-g}\left\{{R\over{2\kappa^{2}}} - {1\over2}
(\partial_{\mu}\phi)^{2}\right\} - \int \sqrt{-g}\, W(\sigma) \, R_{GB}^{2} 
\nonumber \\
     & + S_{\rm mat\!ter}[{\rm matter}\, ;\, g_{\mu\nu}]
\end{eqnarray}

\noindent
This Lagrangian  includes 
the Gauss--Bonnet term coupled to a field  $\phi \equiv \sigma$, for instance 
the modulus field in \cite{antonrizos} where  
$W(\sigma)$ is the coupling of the modulus field to the
Gauss--Bonnet term, $R_{GB}$.

\begin{equation}
R_{GB}^{2} \equiv R^{\mu\nu\rho\sigma}R_{\mu\nu\rho\sigma} -4R^{\mu\nu}
R_{\mu\nu} + R^{2}
\end{equation}

\noindent
The possible coupling of a scalar field 
to the Gauss-Bonnet topological invariant can be constrained  
if one takes into account cosmological supernovae and solar system tests. 
From supernova data gathered in \cite{riess04} and with a parameterized 
analysis of the evolution, it is found that
the present value of the equation of state $w_{0}$ is below --1. 
Though for the measurement of the index of the equation of state 
at $z = 0$ one awaits a large enough nearby supernova sample,  
the results presented in \cite{riess04,riess06} move to examine the
possibility of  $w_{\rm eff} < -1$. 
In the original exploration of the idea \cite{caldwell},
 a Universe which evolves towards 
$w_{\rm eff}$ $<$ --1 would suffer a super--aceleration phase, that has
come to be called the Big Rip.
From an action which includes the Gauss--Bonnet term, one can obtain
$w_{\rm eff} < -1$ for a canonical scalar field (positive 
kinetic energy)  \cite{odintsov,langlois}.  
It avoids the Big Rip for $w_{\rm eff} < -1$. To add a word on this, this
is not the main motivation for Gauss--Bonnet gravity, since  
variations of the braneworld scenario can also get $w_{\rm eff} < -1$ and
avoid the Big Rip \cite{lue,clazkoz,mariam}. The interest on the GB term is 
justified by its presence in a variety of gravity proposals from
string and M--Theory.  Recently some tests using supernovae
have already been done \cite{koiv06}.  The extra GB term opens
naturally many possibilities for the evolution of the scale
factor. The supernova test succeeds for a range of 
effective low energy actions including a Gauss-Bonnet term  \cite{koiv06}.

\subsubsection{Fifth force and equivalence principle constraints}

\noindent
The fields that interact with the dark energy scalar field are constrained
by a large number of tests \cite{uzan03}.
The possibility that the dark energy scalar couples to
nucleons and photons has been investigated  in relation to 
implications for a fifth force  \cite{ellis89,wetter94}.  
Within  current scenarios of particle physics such as MSSM and 
MSUGRA the role of this dark energy  
field shows some observational signatures of the fifth force type as well 
\cite{braxmartin06a}. For instance,
the interaction of the dark energy 
scalar field $\phi$ with the Higgs doublet $H_{\rm u}$ and $H_{\rm d}$
leads to an action where
the  field couples differently to each of
the Higgs $\psi_{{\rm u}}$ and $\psi_{{\rm d}}$ and one expects a violation
of the weak equivalence principle.
In the string scenario, where radion,
dilaton and moduli can be responsible for the early and late
acceleration, one has as well some possible strong couplings of the fields
with matter \cite{damour,neupane,braxmartin06b}. Thus, 
the cosmological context has to be considered together with classical 
gravitational tests on Earth. Thus far, the torsion balance experiments 
set strict limits on the variation of Newton's law  
using different substances and locations. There is a wide range of tests
being done to examine
 the predictions from scalar--tensor proposals and modified
gravity ideas \cite{adelberger06,adelberger03,adelberger05,hoyle04,murphy06}.
 The 
empirical bounds on some types of couplings are tight and worth considering
together with the cosmological probes.

\section{ Cosmological constant?} 

Finding that the negative pressure component gives $w=-1$, would likely be 
the biggest challenge to gravitation theories. At present, SNe Ia have not 
provided such evidence, though averaged values of $w$ are close to --1
at  redshifts up to z$\sim$ 1. To probe that $w$ stays --1 up to
the surface of last scattering, involves refining 
several other methods (growth of density perturbations, 
integrated Sachs--Wolfe effect) \cite{doran}.

\noindent
From the view point of string theorists \cite{witten}, there is no
natural explanation for the vanishing or extreme smallness 
of the vacuum energy. A way to address the small value of the cosmological
constant in the context of a stringy landscape is to consider 
the anthropic principle \cite{susskind}.
However,  invoking the anthropic 
principle as a way out of the problem  entails
 to abandon the quest
for a conventional scientific explanation \cite{witten}. 

\smallskip. 

\noindent
From the observational point of view, the answer is near to 
 $w = -1$, but possible
departures from $w=-1$ are presently being examined through expansions
of the equation of state along the scale factor and other various  
approaches. A expansion often used is \cite{chevpol,linder}: 

\begin{equation}
w(a) = w_{0} + w_{a}(1 -a)
\end{equation}

\noindent
where the scale factor $a = (1 + z)^{-1}$, so $w' = w_{a}/2$ when
evaluated at $z = 1$.

\noindent
 Differences between the 
cosmological constant and departures from it in the equation 
of state are measurable nowadays (see Fig 1). To reach a higher level of 
discrimination between dark energy models
depends on the progress made in the accuracy of the measurements. 
The aim of future projects is to determine distances with 1 $\%$ 
accuracy, i.e. with errors in magnitudes of 0.02 mag \cite{kim04}.
If the equation of state of dark energy is very close to $-1$, 
one can anticipate decades of improving the supernova measurements.
The development that we might witness could be akin to that of the
postnewtonian tests of GR where higher levels of precision are 
being targeted since the middle of the last century into the present.

 \begin{figure}
                                                                     
   \centering
   \includegraphics[width=0.8\columnwidth]{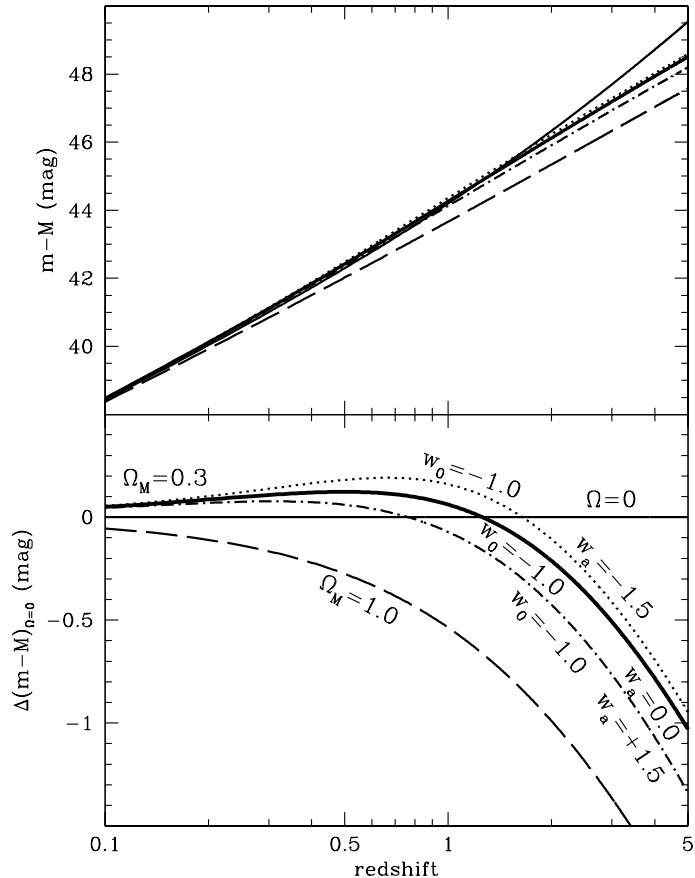}
                                                                               
    \caption{The cosmological constant case (bold line)
    is compared with evolving
   models close to $w=-1$, i.e. a model with $w_{0}= -1.0$ and
 $w_{a} = -1.5$ (short dash line) and a model with $w_{0}= -1.0$ and
  $w_{a} = +1.5$ (dash-dotted line) }

         \label{Figure 1}
\end{figure}

\section{The d$_{L}$ test from SNe Ia} 

\subsection{Overview}

\noindent
A long path has brought us to the present situation where we can already aim
at accuracies of 1$\%$ in distance.
Back in 1968, a first attempt to draw the Hubble
diagram from SNe resulted in
 a spread of $\sigma$ $\sim$ 0.6 mag \cite{kowal}.  Such large scatter was 
due to the inclusion of other types of supernovae, which were later
recognized as being of a different type (SNI b/c) and characterized by
a large scatter in luminosities \cite{filippenko98}.

\noindent
SNe Ia are not standard candles either, but they follow a well known
correlation between maximum brightness and rate of decline of the
light curve. 
In 1977, Pskovskii \cite{pskovskii77} first described the
correlation between the brightness at maximum and the rate of decline 
of the light curve: the brighter SNe Ia have a slower decline of their light 
curves whereas fainter ones are the faster decliners.
A systematic follow--up of SNe Ia
confirmed the brightness--decline relation \cite{phillips93,phillips99}.
The intrinsic variation of SNe Ia is written as a linear relationship:   

\begin{equation}
M_{B} = M_{B0} + \alpha  \ (\Delta m_{15}(B) - \beta) + 5 \ log (H_{0} / 65) 
\end{equation}  

\noindent
$\Delta m_{15}$, the parameter of the SNe Ia light curve family, is the 
number of magnitudes of decline in 15 days after maximum. The value of 
$\alpha$, as well as the dispersion of that relation, have been evaluated in 
samples obtained from 1993 to the present
 \cite{phillips99}. $M_{B0}$ is the absolute magnitude in $\rm{B}$ for a 
template SNIa of $\beta$ rate of decline. An intrinsic 
dispersion of 0.11 mag in that law is found when
  information in more than one color is available.

\noindent
In fact, the 0.11 mag scatter in the brightness--rate of decline
is found when one includes supernovae 
affected by extinction. The scatter is smaller in environments where
dust in the host galaxies of the SNe Ia is a lesser effect \cite{sullivan03}. 
Additionally, the family of SNe Ia form a 
sequence of highly resembling spectra with subtle changes in some spectral 
features correlated with the light curves shapes 
\cite{nugentemp,leib02,matheson05,hook05,lidman,blondin,web}.
 The multiple correlations
allow to control errors within the family of SNe Ia \cite{web}.

\noindent
The parameter $\Delta m_{15}$ requires to have observed the maximum in
the light curve of the supernova. This is not always possible, and 
therefore the various SN collaborations
have formulated the brightness--decline correlation in 
different ways. The High--Z SN Team uses the full shape
of the light curve with respect to a template, method refered as MLCS2k2 
\cite{riess95a,jha}.
 The Supernova Cosmology Project collaboration, on the 
other hand, has introduced the {\it stretch--factor, s,} as a parameter to 
account for the brightness--decline 
relationship \cite{perlmutter98,perlmutter99,goldhaber01}.
The stretch parameter fits the SNIa light curve from premaximum
 to up to 60 days after maximum. Thus, even if the maximum brightness
phase is not observed, the whole light curve provides the decline
value.  More recently, other ways to measure the decline have been 
proposed \cite{tonry,barris04,clocchiatti05}.

\noindent
 Measuring the evolution of the rate of expansion of the Universe 
does not require knowledge of  its present value, $H_{0}$. Large projects have 
been devoted to the determination  of $H_{0}$ by  
attempting the absolute calibration of $M_{B}$ from 
nearby SNe Ia that occur in galaxies with Cepheid--measured distances
 \cite{freedman,sandage}. Controversy persists, though the most recent
values quoted by the two collaborations using this method
are closer (in the range 65--75 km s$^{-1}$ Mpc$^{-1}$) 
\cite{freedman,sandage}.
Though largely ignored as an argument, the mere fact that a SNIa 
is a  explosion of white dwarf  
bounds very efficiently the lower and upper possible values 
of $H_{0}$ \cite{rhubbleneb2}. On the empirical side, the emission at
late phases of SNe Ia gives a precise value of $H_{0}$, since such emission
is very sensitive to the density profile, the mass of the supernova 
and the amount of $^{56}$Ni synthesized in the explosion. Those quantities
are easily measurable from the spectra \cite{rhubbleneb}. The value favored
by this method is H$_{0}$ $=$ 68 $\pm$ 6 (syst) km s$^{-1}$ Mpc$^{-1}$. 

\noindent
This disgresion done, we emphasize the {\it nuisance role}
of the absolute magnitude of SNe Ia when it comes to dark energy cosmology. 
The apparent magnitude of supernovae along z give us the cosmology, 
 $m_{B}$ as \cite{perlmutter99,knop03}:             

\begin{equation}
m_{B} = {\cal M} + 5\, {\rm log}\, {\cal D}_{\cal L}(z;\Omega_{M},
\Omega_{\Lambda}) - \alpha(s-1)
\end{equation}

\noindent
where $s$ is the stretch value for the supernova, ${\cal D}_{\cal L}\equiv
H_{0}d_{L}$ is the ``Hubble-constant-free'' luminosity distance, $d_{L}$ 
is given by (1) and 
${\cal M} \equiv M_{B} - 5\, {\rm log}\, H_{0}
+ 25$ is the ``Hubble-constant-free'' $B$-band peak absolute magnitude of a
$s=1$ SN Ia with true absolute peak magnitude $M_{B}$. Similar use
of the nuisance parameter ${\cal}$ in their light curve fitting 
is made in \cite{riess04,riess06}.

\noindent
In the first determinations of dark energy cosmology, there
were four parameters in the fit: the matter density  parameter $\Omega_{M}$,
the cosmological constant density parameter 
$\Omega_{\Lambda}$, as well as the two nuisance 
parameters, ${\cal M}$ and $\alpha$ (either $\alpha$ itself as used by 
the SCP or a equivalent parameter by other collaborations).
 The four-dimensional ($\Omega_{M}$, 
$\Omega_{\Lambda}$, ${\cal M}$, $\alpha$) space, explored as a grid, 
yields $\chi^2$ and $P\propto e^{-\chi^{2}/2}$ of
the fit of the  
luminosity distance equation to the peak $B$--band magnitudes and redshifts 
of the supernovae. 
After normalization by integrating over the  ``nuisance'' parameters, 
the confidence regions in the $\Omega_{M}$--$\Omega_{\Lambda}$ plane can be
obtained.

\noindent
This use of the magnitude--redshift relation $m(z)$ as a function of 
$\Omega_{M}$ and  $\Omega_{\Lambda}$ with a sample of high--z SNe Ia of 
different z, is a very powerful pointer to the allowed values in the  
$\Omega_{M}$--$\Omega_{\Lambda}$ plane \cite{goobarperl,garnavich}. 
By 1998, such use yielded important results,  
impling that $\Omega_{\Lambda} > 0$  at the 3$\sigma$ confidence level. For a 
flat universe ($\Omega_{T} = 1$), the results from the Supernova Cosmology 
Project meant  
$\Omega_{M}$=$0.28^{+0.09}_{-0.08}$(statistical)$^{+0.05}_{-0.04}$ 
(systematics), 
and the High--Z Supernova Team obtained for a flat universe 
$\Omega_{M}$=0.24$\pm$0.1. The outcoming picture of our universe is that 
about 20--30$\%$ of its density content is in matter and 70--80\% in 
cosmological constant. According to the allowed  $\Omega_{M}$ and 
$\Omega_{\Lambda}$ values, the Universe will expand forever, accelerating its 
rate of expansion.

\subsection{Results on w}

If the size of the SNe Ia sample and its redshift range do allow it, 
one can aim at determining the  average value
of the index of the equation $\left<w\right>$ or, 
going a step further, aim at
determining the confidence regions on the $w_{0}$-- $w_{a}$ plane.
For this later purpose, using the information on the global geometry
as given by WMAP provides a helpful prior. 

\noindent
In Knop et al. (2003) \cite{knop03}, the supernova results were combined with
measurements of $\Omega_{M}$ from galaxy redshift distorsion data and
from the measurement of the distance to the surface of last scattering from
WMAP. The confidence regions provided by the measurement of the
distance to the surface of last scattering  and the SNe Ia confidence 
regions cross in the $\Omega_{M}$--$w$ plane, thus providing very good
complementarity. This work considers an averaged $\left<w\right>$ 
(given the range in
z of the SNe Ia observed). From the fitting assuming $w$ constant, the
value is $\left<w\right>$ 
$=$--1.06$^{+0.14}_{-0.21}$(statistical)$^{+0.08}_{-0.08}$ (identified 
systematics). 

\noindent
The analysis of the SNLS \cite{astier06} also 
 aims at an average
value of $w$, allowed by the z range of this survey. Their 
 71 high--z SNeIa give $\left<w\right>$$=$--1.023 
 $\pm$ 0.090  (statistical) $\pm$
0.054 (identified 
systematics). Recently, the results from ESSENCE, using 60 SNe Ia 
centered 
at z$=$ 0.5 point to a similar result 
with $\left<w\right>$$=$ --1.05$^{+0.13}_{-0-12}$
(statistical) $\pm$ 0.13 (systematics) \cite{wood-vasey07}. In this last 
work, the supernovae from various collaborations are joined with two
separate light curve fitting approaches giving a consistent result for
the average value of $w$.

\noindent
High--z SNe Ia beyond z $>$ 1 enable to have a first look at possible
evolution of $w(z)$. 
The first campaign by Riess et al. (2004) gave the results in the
$w_{0}$--$w'$ plane, using the Taylor expansion for $w(z)$  as
$w(z) = w_{0} + w' z$. The best estimates for $w_{0}$ and $w'$ are 
$w_{0} =$ --1.36$^{+0.12}_{-0.28}$ and $w'=$1.48$^{+0.81}_{-0.98}$. 
Those results \cite{riess04,riess06} exclude a fast evolution 
in the equation of state, and therefore they rule out a number of 
modified gravity proposals. Moreover, the results are confirmed in 
\cite{wood-vasey07} where 
the available SNe Ia at large z,
which can constrain $w_{0}$ and $w_{a}$, are incorporated to test
with a joined sample encompassing a wide range of z.

\noindent
Significant limits have been placed on the evolution of $w(z)$. 
Now the potential of the method relies on an improved
 control of systematic effects.
In this area, there has been substantial progress along the last years.

\subsection{Results on systematic uncertainties}

\noindent
Huge advances have taken place in controlling the uncertainties due to
extinction by dust, the universality of the supernova properties, the
control of lensing by dark matter and the observational process
leading to the the construction of the Hubble diagram of SNe.

\subsubsection{Dust}

The study with the HST of
 the host galaxies of the SNe Ia sample  of 
the Supernova Cosmology Project  \cite{sullivan03}
 allowed to
 subclassify the galaxy hosts of the P99
sample \cite{perlmutter99}. It was found 
 that, at high $z$, early--type galaxies  show a narrower 
dispersion in SNe Ia properties than late--type galaxies, as they do at low 
$z$ (\cite{branchbaron06}).

\noindent   
This result is very encouraging as it supersedes previous dispersion
values obtained in samples of supernovae in all galaxy types. 
Supernovae in  dust--rich environments, such as spiral galaxies, 
are more affected by extinction. The supernova
magnitudes are corrected regularly from Galactic extinction
\cite{schlegel}, but
the correction by dust residing in the host galaxy or along the line of sight
is often 
not included in the standard fit, as it would require to have extensive 
color information that might not be available for high--z SNe Ia.
When it is included, it is still 
subject to the uncertainty in the extinction law \cite{jha}.

\noindent
 Let us, for instance, consider the correction of dimming by dust as entering 
 in the term $A_{B}$, the magnitudes of extinction in the $B$ band.
 We take, for instance, a supernova
 of a given stretch s, and we place its $m_{\rm eff}$ (the magnitude of
 the equivalent supernova of $s=1$) 
 in the Hubble diagram. We might use, as in \cite{perlmutter99,knop03}: 

\begin{equation}
m_{\rm eff} = m_{B} + \alpha(s - 1) - A_{B}
\end{equation}

\noindent
The extinction in the canonical band $B$ 
 of the spectrum ($A_{B}$) is determined by 
the extinction coefficient $R_{B}$ in that band. Any uncertainty 
in this coefficient as due
to variation of dust properties, leads to uncertain  estimates  of 
$A_{B}$. $R_{B}$ is of the order of 4.1 $\pm$ 0.5 as properties of dust
might change going to high z. 
 If we have an error of 
0.1 in the observable $E(B-V)$ due to dust extinction, it gets amplified as:

\begin{equation}
A_{B} \equiv R_{B} \times E(B - V) 
\end{equation}

\noindent
The present way of dealing with SNe Ia, including information 
in several photometric bands of the spectrum where extinction
coefficients ($R_{V}$, $R_{I}$) are lower and one gets smaller 
values in the corresponding  $A_{V}$ and $A_{I}$, allows to control
the dust problem. In fact, the coefficients $R_{B}$, $R_{V}$ and $R_{I}$
are also determined by methods determining distances to SNe Ia.
In a sample of 133 nearby SNe Ia, it is found in \cite{jha}, that the dust 
in their host galaxies is  well described by a mean extinction law 
with $R_{V}$ $\simeq$ 2.7. 
In all this, the level at which SNe Ia are treated, including how many 
bands are used, becomes critical. 
 A  use of SNe Ia without taking into account color information
leads to a dispersion of 0.17mag. When taking into account color
 information, the limit goes down to 0.11mag. Limiting the effect of 
 extinction, as in SNeIa in ellipticals should bring down the dispersion.
 This finding is motivating
 the ongoing SN searches in clusters of elliptical galaxies 
\cite{perlmutter05}. In environments where dust is controlled, 
 such as SNe in clusters
rich in elliptical galaxies,
 the systematic effect introduced by dust extinction
can be better controlled. There, one can also control the evolution of 
dust--properties.

\noindent
In Table I, the level of accuracy in the first uses of SNe Ia back in 
 1998 is shown, with the contributions of each systematic and statistical 
error. The improvement is given for a future mission able to control
the systematics in the best possible way. It is 
foreseen that several sources of systematic errors will become negligible  
within the next decade.

\subsubsection{Evolution or Population drift}

This question resides on whether we know that SNe Ia properties evolve 
with z. Much has been done along this path. Nowadays, while we know that
 dust is the unavoidable systematics, several studies 
confirm the universality of the light curve properties of SNe Ia. 
The study of a significant sample of SNe Ia at 
various $z$ by the Supernova Cosmology Project reveals that the rise times to 
maximum of  high--z SNe Ia are similar to the low--$z$ 
SNe Ia  (\cite{aldering00,goldhaber01}).
 The same is found in the SNLS sample (\cite{conley06}). 
 Statistical evaluation is so far consistent 
with no difference between the low--$z$ and high--$z$ samples.

\noindent
Moreover, the possibility of existence of an extra parameter in the maximum 
brightness--rate of decline relation has been carefully examined. Among 
possible influences, metallicity was one of the obvious ones to consider. 
Examining supernovae in galaxies with a gradient of 
metallicity, it is found  \cite{ivanov00}
 that there is no evidence for metallicity 
dependence as an extra parameter in the light curve correlation of SNe Ia. 
This result comes from an analysis of the SNe Ia light curves of
 a sample of 62 supernovae in the local Universe. The SNe Ia
 belong to different  
 populations along a metallicity gradient. 
This check has been done as well at high--z 
\cite{quimby02} using
74 SNe Ia (0.17 $<$ z $<$ 0.86) from the SCP sample. No significant 
correlation between peak SNIa luminosity and metallicity is found.

\noindent
Often we address the question of the spread of the SNe Ia 
samples in galaxies of different types: on how many fast SNe Ia versus 
intermediate or slow decliners are found in the various morphological types. 
Spirals at low or moderate redshifts should encompass all ranges of variation 
in the SNe Ia properties since they contain populations with a wide spread in 
age. The cosmological SN collaborations find that in those samples the 
one--parameter correlation does give a good description of the variation of 
the SNe Ia light curves: there is no residual correlation after 
the light--curve shape correction 
\cite{goldhaber01}.
   
\noindent
One can go a step further and test the SNIa objects themselves. 
This has been attempted in a intermediate z sample and it is a feasible
project at all z \cite{web}.  The physical diagrams of SNe Ia 
at intermediate z and at low z seem to be similar \cite{web}. Those diagrams
reveal the composition, velocity gradient and radiation properties of SNeIa 
\cite{web}. Looking 
at the inside of the SNe Ia is the best way to test whether they are 
the same at all z. Such step forward is giving definitive results
which reaffirm the validity of the SNe Ia method.
 The progress done in this domain is summarized in Table I.

\subsubsection{K--corrections} 
 
There is another element
 yet to be introduced here: it deals with the K correction.
The K--correction is necessary to calculate the apparent magnitude in a ${\it
y}$ filter band of a source observed in a ${\it x}$ filter. As we are 
comparing the observed flux of supernovae in a given spectral band
along the expansion history, we need to take into account the effect that
the photons of the supernova have shifted to redder wavelengths and spread 
over a different spectral wavelength range and the  
 observed flux is different from  that at emission. 

\begin{equation}
m_{Y}(z) = m_{X}(z = 0) + K_{xy}(z)
\end{equation}

\noindent
where

\begin{equation}
K_{xy} =  2.5\; {\rm log}\, \left\{(1 + z){{\int F(\lambda)S_{x}(\lambda)
d\lambda}\over{\int F\left[\lambda/(1 + z)\right]S_{y}(\lambda)
d\lambda}}\right\} 
+ {\cal Z}_{y} - {\cal Z}_{x}
\end{equation}

\noindent
where $F(\lambda)$ is the source spectral energy distribution (a SN 
in this case), and $S_{x} (\lambda)$ is the transmission of the filter x.
The ${\cal Z}_{y} - {\cal Z}_{x}$ term accounts for the different zero 
points of the filters \cite{kim96}. At present, this term amounts to
an uncertainty of 0.02mag in the overall budget of SNeIa as distance 
indicators, but it can be decreased to $0.01$ mag. 

\noindent
The comparisons of  supernovae 
with the fine time sequences of spectra along all 
phases  available, are turning the uncertainty in the
 K--correction into a decreasingly small figure.  
The large amount of spectral
data  on SNe Ia, and the good correlation with photometric
properties, can indeed make this uncertainty very small.

\subsubsection{Gravitational lensing of SNe Ia}

\noindent
Gravitational lensing causes  dispersion in the Hubble diagram for 
high redshift sources. There was an early concern on the systematics
and bias that this effect could cause in the determination of the
cosmological parameters using SNeIa \cite{amanullah03}. 
It has been shown through the study of lensing in the
highest z SNe Ia sample \cite{jonsson06} that the magnification distribution
of SNeIa matches very well the expectations for an unbiased sample, i.e., 
their mean magnification factor is consistent with unity. The effect can 
be very well controlled by studying the galaxy fields \cite{jonsson06}. 
 
\noindent
Moreover,
 SNe Ia should show  a negligigle systematics caused by gravitational lensing
when having a few SNeIa per redshift bin \cite{wood-vasey07}. This means that
in future missions that effect should not have impact in limiting the
accuracy of $w(z)$. 

\subsubsection{ The limit of SNeIa as calibrated candles}

\noindent
 The better
 we deal with the light curve and spectral information of SNe Ia, 
 the lower becomes the 
 dispersion in the Hubble diagram. If 
 SNe Ia have 
 a systematic error in their use (independently from evolution),
 this is likely well below the accuracy to which we would like to determine
 distances (1$\%$). We have not yet hit that brick wall. 
 Using 
 multi--epoch spectral information and accurate light curves in various
 filters should enable to test the power of these candles and their 
 limiting irreduceable error \cite{web}.

\subsubsection{A joined sample}

 Supernovae collected along years show the effect of observations 
 treated with various degrees of accuracy. The dispersion 
 around the best fit reflects that fact.
 For instance, some samples would have  
 information on only two colors and at very few epochs, 
 making it difficult to control reddening, or supernovae had 
 no spectrum confirming that they were SNe Ia. The new samples 
 will naturally have a higher degree of accuracy in magnitude and
 extinction for each supernova than previous samples. The ones with
 poorer information  have been placed in \cite{riess04,riess06}
 into the ``silver´´  category (as opposed to ``gold'' category).
 However, even within the Gold sample the requirements have changed: 
 the latest definition of Gold sample demands higher quality standards than
 in \cite{riess04}. In some of these SNe Ia lists, distance moduli
 of supernovae are assembled but they have been obtained through
  procedures which 
 deal differently with dust extinction. The Hubble diagram resulting from
 those compilations would naturally show a large dispersion. 
 In \cite{necesseris}, it is argued that a few supernovae of an early  
 High Z SN Search sample are pulling $w(z)$ towards evolution. The exploration
 done with a method independent from that in \cite{riess04} showed that
 the central value in the 1$\sigma$ confidence varied according to the 
 light curve and reddening fitter of those SNe Ia \cite{rvalencia}. 
 Fortunately, supernovae enable to compare results obtained 
 with different light curve
 fitting approaches. The published photometry by the various collaborations
 can be used to bring together
  the SNe Ia by using a consistent procedure into the 
 Hubble diagram. This was done for the SNe Ia considering available up
 to 2005 in \cite{rvalencia}. 
  This is done in \cite{wood-vasey07,rubin,kowalski} for
 samples including the latest 
 data from the Higher Z collaboration.

\section{Testing the adequacy of a FRW metric }

 In most of the above discussion on dark energy, the FRW metric 
 is taken as the right metric for our Universe. Modern cosmology 
 supports that the Universe is highly isotropic on average on high
 scales. The main evidence is coming from the statistical isotropy of
 the Cosmic Microwave Background radiation. There are also
 cosmological observations supporting a matter distribution homogeneous
 on large scales, of the order of 100 h$^{-1}$ Mpc.  Given
 the observed structures of matter in voids and filaments, it is worth
 considering whether departures from the homogeneous and isotropic
 Universe could alter the present discussion, in particular, the possibility
 that there were 
 no need for dark energy, but instead, that the observations are the result 
 of an effect of inhomogeneity \cite{blarena}.

\noindent
  As pointed out in \cite{bonvin06},
large samples of supernovae will soon allow  to test the luminosity distance
relation in various directions and redshift bins. One can consider
the luminosity distance d$_{L}$ as a function of direction
${\bf n}$ and redshift z \cite{bonvin06}.   

\smallskip

\noindent
The direction--averaged luminosity distance is: 

\begin{equation}
d_{L}^{(0)}(z) = {1\over{4\pi}}\int d\Omega_{\bf n}d_{L}(z,{\bf n}) = (1 + z)
\int_{0}^{z}{dz'\over{H(z')}}
\end{equation}

\noindent
This is the usual value retrieved by the SN collaborations. 
But its directional dependence $d_{L}(z,{\bf n})$ can be a test of the
validity of the isotropy assumption. The authors in \cite{bonvin06} 
propose to expand the
luminosity distance in terms of spherical harmonics, to obtain the 
observable multipoles $C_{l}(z)$. 
This idea was also investigated in \cite{riess95a}, where 
 the motion of our Local Group
with respect to the CMB was measured. Results in \cite{bonvin06} are
 in agreement with the results in 
\cite{riess95a}.

\noindent
Moreover, we can think of using the dispersion in the magnitude of SNe Ia
in each direction ${\bf n}$ to put constrains on anisotropic models. 

\noindent
As discussed in \cite{conference}, the dispersion of the magnitude in the
Hubble diagram is at odds with some simple inhomogeneous models that
predict a higher spread in $m(z)$ and also a non--linear
behavior.  

\noindent
The anisotropic and inhomogeneous
models have to confront present and future values on the SNeIa
dispersion at each redshift interval $z_{bin}$ and in every  direction
 ${\bf n}$:

\begin{equation}
{{\sigma({ d_{L}(z_{bin}, {\rm\bf n})}})\over{d_{L}(z_{bin}, {\rm\bf n})}} = 
{{{\rm ln}(10)}\over5} \sigma({ m(z_{bin}, {\rm\bf n})})
\end{equation}

\noindent
 The scaling 
 solutions of the regionally averaged cosmologies can simulate the
 presence of a quintessence field or other negative pressure
 component \cite{blarena,celerier00,kriotto}. But, any backreaction
 from inhomogeneity and anisotropy would show in the dispersion
 in the Hubble diagram. Those challenges have to be met by the non--FRW
 proposals.

\noindent
 Up to now, the exploration of toy models calculating the effect of
 inhomogenities in the scale factor, find irregularities in
 the expansion rate beyond what is observed in the SN Ia Hubble diagram.
 The dispersion in the Hubble diagram of a standard candle is
 increased \cite{conference}.

\noindent
 The situation concerning the predictions of inhomogeneous models 
 \cite{blarena, rasanen, kriotto} is summarized
 in \cite{celerier00}. It is concluded that the
 inhomogeneities that would be able to amount to a $d_{L} (z)$ effect 
 as observed from supernovae must be subhorizon and of strong type and
 they require a treatment beyond first--order perturbation theory. 
 Research in this field will continue towards  quantitative 
 predictions to improve the connection with observational
 cosmology \cite{buchert}.

\section{Next decade experiments} 

The aim of discriminating between various possibilities for dark energy
places stringent requirements on the use of standard candles \cite{maor}. 
Within the two
decades from the discovery (1998) and the gathering of space
mission data, it is feasible to reduce largely the errors, as mentioned
before, and this would reflect into an error of only 
 1$\%$-- 2$\%$  in $d_{L}(z)$.

\noindent
Errors much bigger than this 1$\%$ of error in d$_{L}$ result in 
the impossibility to distinguish between models. Even at the level
of 1$\%$ in d$_{L} (z)$  one finds some degeneracy \cite{maor,espana}. 
 Space projects aimed at determining the nature of dark energy 
 are counting on a systematic error in the method around 
 0.02 mag (i.e. 1\% in the distance) \cite{kim04}.
 
\noindent
 A different question is whether from ground, within the 
planned experiments at work in the present decade, one can converge
towards that limit. 

\noindent
These questions have been investigated in detail by using Monte Carlo
simulations including the systematic errors in $m(z)$ (or $d_{L}(z)$)
to be found in the evaluation of  
the present value of the equation of
state of dark energy $w_{0}$ and its first derivative in the 
scale factor $w_{a}$  \cite{kim04} (see expression 33).

\noindent
SNAP is a mission that has been specifically designed to 
gather about 2000 SNe Ia up to z $\sim$ 1.7. As a single mission 
gathering supernova data with the same instrumentation, it can eliminate
possible photometric offsets and include high--precision photometry
out to 1.7 microns\cite{snapweb}.

\noindent
 With such large number of SNe Ia, 
 statistical errors in every redshift bin of $\Delta z$ $=$ 0.1 become
 negligible. The error is dominated by the systematics. 
 The total error in each redshift bin is given by:

\begin{equation}
\sigma_{bin} = \sqrt{{{\sum\sigma_{st}^{2}}\over{N_{bin}}} + \sigma_{sys}^{2}}
\end{equation}

\noindent
with expected systematic error;

\begin{equation}
\sigma_{sys} = 0.02\,\, z/z_{max}
\end{equation}

\noindent
 Table I gives a progress evaluation in the control of 
 high--redshift supernova
distance uncertainties.

\begin{table*}[hb]
\footnotesize
\begin{center}
TABLE I \\
\vskip 9pt
ERROR CONTRIBUTIONS TO HIGH SUPERNOVA DISTANCES in 1998 \\
\vskip 7pt
\begin{tabular}{lc|lc}
\hline
\hline
Systematic Uncertainties ($1\sigma$) & Magnitude & Statistical Uncertainties 
($1\sigma$) & Magnitude \\
\hline
Photometric System Zero Point & 0.05 &  Zero Points, S/N   & 0.17 \\ 
Evolution & \hskip -12pt $<0.17$        & K-corrections & 0.03 \\
Evolution of Extinction Law   & 0.02    & Extinction            & 0.10 \\
Gravitational Lensing         & 0.02    & Light curve sampling & 0.15 \\
\hline 
\end{tabular}
\end{center}
\end{table*}

\vskip -10pt

\begin{table*}[hb]
\footnotesize
\begin{center}
PROSPECTS TOWARDS 1\% ERROR IN $d_{L}(z)$ in  a space mission \\
\vskip 3pt
\begin{tabular}{lc|lc}
\hline
\hline
Systematic Uncertainties ($1\sigma$) & Magnitude & Statistical Uncertainties 
($1\sigma$) & Magnitude \\
\hline
Photometric System Zero Point$^{a}$ & 0.01 & Zero Points, S/N  & 0.00* \\ 
Evolution                     & 0.00*    & K-corrections       & 0.00*
\\
Evolution of Extinction Law   & $< 0.02$    & Extinction          & 0.00* \\
Gravitational Lensing &  $<0.01$ & Light curve sampling
  & 0.00* \\
\hline 
\end{tabular}
\end{center}
\vskip -3pt
\hskip 1 true cm $^{a}$ 0.00* means values of the order of 10$^{-3}$.
 The statistical errors are for the redshift bin. 

\hskip 1 true cm $^{b}$ In the Table for prospects in a space mission, 
the statistical error from light curve sampling 

\hskip 1 true cm of SNe Ia is reduced due to the large number
of SNe Ia data per redshift bin.

\end{table*}

\noindent
Ground--based projects will not be able to achieve such accuracy. They 
aim at a reduction of systematic errors to 0.04--0.05 mag. 
Much of what will be feasible from now until the next space era 
of determination of the equation of state, depends on the use of 
all the information already gathered in the best way. 
Spectra gathered so far can be exploited further than what it has 
been done up to now, in order to reduce systematic errors.

\noindent
In the coming years, a lot can be done to improve the present
dispersion of the Hubble diagram of SNe Ia and aim at
a scatter of 1$\%$ in d$_{L}$. The above limit is only reachable if 
all the systematic effects are dealt with to the best.

\section{Complementarity}

\noindent
The SNe Ia test of $w(z)$ benefits from  using information on 
$\Omega_{M}$ obtained by some other method.
 This can come from weak lensing or from large scale 
structure \cite{heavens,tegmark,percival}.
Baryon acoustic oscillations can also provide a measurement of 
distance along z, in this case, angular distances $d_{A}(z)$ and 
of $E(z) = H(z)/H_{0}$ \cite{eisenstein05,nichol06,yamamoto}.

\noindent
Generally, for any geometry, one has:

\begin{equation}
A = {{\sqrt{\Omega_{M}}}\over{E(z)^{{1\over3}}}}\,\left[{1\over{z 
\sqrt{|\Omega_{K}|}}} S\left(\sqrt{|\Omega_{K}|}
\int_{0}^{z}{dz'\over{E(z')}}\right)\right]^{{2\over3}}
\end{equation}

\noindent
A measurement of $d_{A}$ 
has been obtained \cite{eisenstein05}.

\begin{equation}
A = {{\sqrt{\Omega_{M}}}\over{E(z_{1})^{1/3}}}\,\left[{1\over{z_{1}}} 
\int_{0}^{z_{1}}{dz\over{E(z)}}\right]^{2/3}\; =\; 0.469\pm0.017 
\end{equation}

\noindent
where  $z_{1} = 0.35$ is the redshift at which
the acoustic scale has been measured in the redshift sample.

\noindent
Combining BAO and SNeIa distances one can aim at tracing dark energy
all through a high z range.
 The value of $w(z=z_{rec})$ is provided
by WMAP and Planck. In between, other methods such as studying the growth 
of large scale structure (LSS)
 obtained from Ly$\alpha$ clouds and clusters  would depict 
the growth factor of structure along z. 
Considering the information at high z derived from LSS, SNe Ia and
CMB should lead to learn about early dark energy  
\cite{frieman02,freese,doran,tegmark,
dunkley,kunz06}.

\section{Conclusions}

\noindent
Type Ia supernovae can test GR against other alternative theories of 
gravitation. 
Theories of higher dimensional gravity have different 
Friedmann equations from those originating from GR. Some of those
modified gravity proposals have already been tested and ruled out
through the Hubble diagram of SNeIa up to very high z.

\noindent 
If dark energy is due to a scalar field either
minimally coupled to gravity (i.e. quintessence) or non--minimally
coupled to gravity (arising from scalar--tensor gravity or the effective 
low--energy superstring proposals), the Hubble diagram up to z $>$ 1 
obtained by the supernova collaborations gives the opportunity 
for exhaustive tests. 
In what concerns scalar--tensor gravity,  supernovae are 
a precision experiment in a way analogous
to the observations of the PSR B1913$+$16 or to solar system tests.

\noindent
Moreover, supernovae have already provided hints on the form 
of the potential of the hypothetical scalar field
(for the case of minimally coupled field) that could be associated 
to the acceleration of the Universe. 

\smallskip

\noindent
Nonetheless,
GR plus a form of cosmological constant is not ruled out observationally.  
Finding out how is  $w$ close to $z=0$ is one of the most intriguing
questions. Hopefully, this will be clarified by the samples to come. The 
goal of obtaining a consistent
 picture on $w(z)$ up tp z $\sim$ 1.5 from the present samples is approached
 through the various SN procedures used by the existing collaborations.

\smallskip

\noindent
Ultimately, supernovae together with complementary methods just recently 
discovered shall select the correct ideas on dark energy and
gravity. 
If any modified gravity theory has a cosmological effect, those
methods should start to see it. 

\bigskip

\noindent{\bf Acknowledgements}

\noindent
I would like to thank  Alex G. Kim and Marek Kowalski for reading the 
manuscript and pointing to useful suggestions. This work has been 
supported by grant AYA2006--05369.

\vfill\eject

\end{document}